\def\np{Nucl. Phys. }
\def\pl{Phys. Lett. }
\def\prl{Phys. Rev. Lett. }
\def\pr{Phys. Rev. }

\def\be{\begin{equation}}
\def\ee{\end{equation}}
\def\ba{\begin{eqnarray}}
\def\ea{\end{eqnarray}}

\documentstyle[12pt]{article}
\begin{document}

\title{{\bf An algorithm to compute Born scattering amplitudes without
Feynman graphs}}

\author{{\bf Francesco Caravaglios}\thanks{INFN Fellow},
\\ {\it Department of Physics,
Theoretical Physics,
University of Oxford,}\\
{ \it 1 Keble Road,
Oxford OX1 3NP} \\ \it (e-mail: caravagl@thphys.ox.ac.uk) \\
 \\
{\bf Mauro Moretti,}\thanks{European Community Fellow,
CHRX-CT93-0132 contract, Human Capital and Mobility Program}\\
{\it Department of Physics, The University,} \\
{\it Southampton, SO17 1BJ, UK}
\\ \it (e-mail: moretti@hep1.phys.soton.ac.uk)}
\date{}
\maketitle

\begin{abstract}
In this paper we suggest an {\it iterative} algorithm
 to compute automatically  the scattering matrix elements
of  any given effective
lagrangian, $\Gamma$. By exploiting the relation between
$\Gamma$
 and the connected Green function generator, $Z$, we provide a formula
which   does not require the use of the
Feynman graphs and it is suitable to implement a numerical routine.
By means of this algorithm  we have built a relatively simple and fast fortran
code which  we have used to calculate, at the tree level,
 the rate of
four fermion
production at LEP I\negthinspace{I} (finding a very good agreement
with previous calculation) with and without the emission of
one observable photon.
\end{abstract}
\vspace{-16cm}\rightline{ OUTP 9528P}
\rightline{SHEP 95-22}
\vspace{15cm}

\subsection*{Introduction}

The computation of  the  matrix element of physical processes is often a
difficult task.
With the increasing collider energies the complexity of the required
computations has grown considerably.
When the number of final particles is high,
even to evaluate the corresponding tree level Feynman
 diagrams
becomes hard and the final formula is often
an intricated  function of several variables,
inadequate for practical use.

For this reason, many authors  made a lot of work to simplify
these calculations. Very good achievements have been obtained:
the helicity amplitude technique \cite{elicity} shows how to manipulate
the Feynman rules to compute by hand some extremely difficult
multi-particle processes
which otherwise would be impossible to handle,
color decomposition, recursion relations \cite{color} and
 string inspired \cite{string} rearrangement of
the diagrammatic expansion have allowed a drastic simplification
in the number of graphs to be considered in some QCD processes,
programs of symbolic manipulation provide a very useful tool to
manage and evaluate the Feynman graphs \cite{symbolic}
and finally a completely automatic treatment of the Feynman rules
is available \cite{kaneko}.

  Here we pursue a  different approach.
 By exploiting the relation among the
one particle irreducible Green Functions generator $\Gamma$ with
 the connected Green Functions generator $Z$ we are able to provide
an explicit formula for $Z$ (for a given $\Gamma$) as a series expansion
and we show that after the truncation of the series to a proper
 number of steps the exact scattering amplitude
is recovered. We use this formula to implement a FORTRAN code
and we present the computation of the  rates for the processes
$e^+e^-\rightarrow \mu^+\nu_\mu\tau^-\bar\nu_\tau$
and
$e^+e^-\rightarrow \mu^+\nu_\mu\tau^-\bar\nu_\tau
\gamma$ for  some center of
mass  energies  accessible at LEP I\negthinspace{I}.

\subsection*{An iterative algorithm for the scattering matrix}
\
If the one particle irreducible Green Functions generator, $\Gamma$,
 of a theory is
known
the computation of the S-matrix requires the evaluation
of  the  legendre transform, Z, of $\Gamma$ \cite{zuber}
\be
Z(j_1,\dots , j_n)= - \Gamma(\varphi_1,\dots,\varphi_n)
+ j_k(x)\varphi_k(x)
\label {legtr}
\ee
where $\varphi_j$ are the classical fields defined as the solution of
\be
 J_k =\frac  {\delta \Gamma }{\delta \varphi_k }
\label{minim}
\ee
and the $J_k$
play the role of classical sources.

For the sake of clarity, in the following we will stick to
a specific example, namely we discuss
the process
\be
\label{process}
\phi(q_1)+\phi(q_2) \rightarrow \phi(q_3)+\phi(q_4)
\ee
in a $\lambda \phi^3$ theory.
As a matter of convenience we use the convention that all the particles
 are incoming and therefore the momenta of the outcoming particle take an
overall minus sign.
  From now on we define
\be
p_1=q_1~~p_2=q_2~~p_3=-q_3~~p_4=-q_4
\ee
 The lagrangian\footnote{
In the following example we take the most simple lagrangian, but the
generalization to an arbitrary effective lagrangian (see after) which also
includes fermions and higher order operators (e.g. radiative
corrections terms calculated by hand) is almost straightforward.}
              ${\cal L}$ of the model, is
\ba
{\cal L}&=&\int{d^4x~ (- \frac{1}{2}
\phi\partial^2\phi + \frac{\lambda }{6}\phi^3)}
\nonumber \\
&=&\int{{d^4p\over (2 \pi)^4}
 d^4q ~\frac{1}{2}p^2 \delta(p+q)\phi(p)\phi(q)  }
+\frac{\lambda }{6}
\int{{d^4p\over (2 \pi)^4}{ d^4q \over (2 \pi)^4}
d^4k ~\delta(p+q+k)\phi(p)\phi(q)\phi(k)}.
\nonumber \\ \label{example}
\ea
Hence  the
Z functional for the theory  is, at the tree level, the
 Legendre transform of (\ref{example})
\be
Z=-{\cal L} + \int d^4 x ~ \phi(x) J(x)
\label {zl}
\ee

In order to calculate the scattering matrix element for the process
(\ref{process})
we specify the source $J(x)$ to be\footnote{
Even if theoretically  one should take only the real part of such
$J(x)$, in practice, for our pragmatic aim   this is
not significant.}
\be
J(x)=\sum_{l=1}^4 a_l e^{i p_l x}
\label{source}
\ee
where $a_l$ is a
     coefficient which might carry
lorentz and internal symmetry indices.
The required matrix element is obtained
after differentiation of $Z$ with
respect to the $a_l$ in the limit
$a_l\rightarrow 0$ and after having properly cut
the external legs.
With the source $J(x)$ of eq.~(\ref{source}) the functional $Z$ becomes
\be
\label{matrix}
Z(a_1,a_2,a_3,a_4)=-{\cal L} +\int{d^4 x \sum_{m=1,4} a_m e^{i p_m \cdot x}
\phi(x)}
\ee
where
  $\phi(x)$ is such that it
 minimizes  (or more generally it picks up a stationary point)
  the right hand side of eq.~(\ref{zl}) for any fixed set of $a_m$,
{\it i.e.}   it is of the form
\be
\phi(x)= b_m e^{-i P_m x}
\label{phi}
\ee
 with
\be
P_m = c_m^l p_l  \qquad c_m^l=0,+1.
\label{mom}
\ee
In fact, even if the the fulfillment of the
stationary condition for an arbitrary choice   of
the $a_j$  would require $c_m^l$ to take any
integer value, we are
interested in the Green functions which are recovered only after
taking the limit $a_j\rightarrow 0$. In such a limit, these are the
only momenta which survive and which contribute to the final result\footnote{
After, when the explicit iterative solution
for the stationary condition will be given, it will be manifest that
other momenta will show up only with higher order $a_j$.}.
In our specific case
 we have fourteen $P_j$ (and $b_j$)

\begin{center}
\begin{tabular}{l l}
$P_1=p_1$;& $P_2=p_2$; \\ $P_3=p_3$;& $ P_4=p_4$; \\ $P_5=p_1+p_2$; &
$P_6=p_1+p_3$;\\ ...  & ... \\ $P_{13}=p_1+p_3+p_4$;
& $P_{14}=p_2+p_3+p_4$;\\
\end{tabular}
\label{expmomenta}
\end{center}

Replacing $\phi(x)$ in eq.~(\ref{matrix}) we obtain, in momentum space,
\be
Z(a_1,a_2,a_3,a_4)
 = - \frac{1}{2}
\sum_{j,l}b_jb_l  \Delta_{jl}-\frac{1}{6}\sum_{j,l,m}
b_j b_l b_m D_{jlm}
+
\sum_{m,l}a_m b_l \frac {\Delta_{ml}} {P_m^2} \label{zexample}
\ee
where the $D_{ijk}$ and $\Delta_{lm}$
take into account the momentum conservation, i.e.
\ba
D_{ijk}&=&\lambda ~~~~~~~~~~~~~if~~ P_i+P_j+P_k=0 \nonumber \\
D_{ijk}&=&0 ~~~~~~~~~~~~~if~~ P_i+P_j+P_k\ne0. \nonumber \\
\Delta_{ml}&=&P_m^2~~~~~~~~~~~~~if ~~P_m+P_l=0 \nonumber \\
\Delta_{ml}&=&0~~~~~~~~~~~~~if ~~P_m+P_l\ne 0. \\
\nonumber \ea
and the $b_j$ variables in eq.~(\ref{zexample}) satisfy the minimum equations
\ba
 a_i&=&{d {\cal L} \over d b_i}
= \Delta_{il} b_l +{1\over 2} D_{ijk} b_j b_k ~~~~~~i=1,4
\nonumber \\
&and&  \nonumber \\
 0&=&{d {\cal L} \over d b_i}=  \Delta_{il} b_l
 +{1\over 2} D_{ijk} b_j b_k
{}~~~~         i>4 \label{eqsol}\\
\nonumber
\ea
Here an important point should  be noticed:  we began to study the problem
of finding the functional $Z$ of eq.~(\ref{zl}) which is, formally,
a problem with an infinite number of degrees of freedom. Using
eq.~(\ref{source}) with the constraint given in eq.~(\ref{mom}) we have
 reduced it to the solution of eqs.~(\ref{eqsol}) where the number of degrees
of freedom is finite.

We now solve  iteratively  eq.~({\ref{eqsol})
by replacing the $b_j$ with the
 sum
$b_j=\sum_{r=0}^\infty b_{j,r}$
where the $b_{j,r}$ are iteratively defined as functions of the $b_{j,s}$,
with $s<r$,
as follows:
\ba
1 &=& b_{i,0}~~~~~~~~i=1,4 \nonumber \\
0&=&b_{i,0}~~~~~~~~i>4 \label{s1}\\ \nonumber
\ea
where we set $a_i=p_i^2$ to obtain the truncated Green function.
At the second step the same eq.~(\ref{eqsol}) gives
\be
b_{j,1}=-{ 1\over 2 }\Delta^{-1}_{j,l}  D_{lkm} b_{k,0} b_{m,0}
\label{s2}
\ee
and finally as  last step
\be
b_{j,2}=-{1 \over 2 }\Delta^{-1}_{j,l}  D_{lkm}
(b_{k,0} b_{m,1}+b_{k,1} b_{m,0})
\label{s3}
\ee
To converge to the exact solution of eq.~(\ref{eqsol}) one should
continue the series of the $b_{j,r}$ up to $r=\infty$, however, for the
purpose of computing the scattering amplitude, the terms we evaluated in
(\ref{s1},\ref{s2},\ref{s3}) are sufficient. In fact the $b_{j,r}$
series can be  regarded as an expansion in the coupling
$\lambda$  and we have indeed truncated the series to the order
$\lambda^2$.

We now substitute the obtained $b_{j,r}$
 in eq.~(\ref{zexample}).
The result
  coincides with the required scattering amplitude
\ba
{\cal A}_{p_1p_2p_3p_4}&=& p_1^2 p_2^2 p_3^2 p_4^2
\lim_{a_j \rightarrow 0} \frac
{\partial^4 Z(a_1,a_2,a_3,a_4) }
{\partial a_1 \partial {a_2}\partial {a_3}\partial {a_4} }
\nonumber \\
&=&-{1\over 2}\sum_{s+r=2}\sum_{j,l=1}^{14} b_{j,r}b_{l,s}
\Delta_{jl}-{1\over 6}
\sum_{s+r+t=1}\sum_{j,l,m=1}^{14} D_{jlm}b_{j,r}b_{l,s}b_{m,t}
\nonumber \\ & & \quad
+ \sum_{l=1}^4\sum_{j=1}^{14}b_{j,2} \Delta_{jl} \label{intermedio}\\
&=&\lambda^2({1\over (p_1+p_2)^2}
+{1\over (p_1+p_3)^2}+{1\over (p_1+p_4)^2}). \\ \nonumber
\ea
One must notice that
eq.~(\ref{intermedio}) contains some redundacy and if one is
wishing to use it to build a numerical routine he would immediately
face the problem to handle  quantities of the type
$0/0$.
It is however possible to use eqs.~(\ref{eqsol}) to simplify
eq.~(\ref{intermedio}) (and simultaneously overcoming the problem
of dealing with $0/0$ expressions).
Multiplying both sides of eq.~(\ref{eqsol}) by $b_i$ and summing over
the index $i$, one obtains, from eq.~(\ref{intermedio}),
\be
p_1^2 p_2^2 p_3^2 p_4^2
\frac {\partial^4 Z(a_1,a_2,a_3,a_4) }
{\partial a_1 \partial {a_2}\partial {a_3}\partial {a_4} }
=
-{1\over 6}
\sum_{s,r,t}\sum_{j,l,m=1}^7 D_{jlm}b_{j,r}b_{l,s}b_{m,t}
\label{simple}
\ee
where the indices $j,l,m$ are  defined according to the conventional
ordering given above.
 Notice that the evaluation of eq.~(\ref{simple}) does not require
the knowledge of the $b_{j,2}$ of eq.~(\ref{s3}) which are non zero only
if $j\ge 10$. Due to this fact one does not need to evaluate the
inverse of the propagator for on shell four momenta which occurs in
eq.~(\ref{s3}).

For an arbitrary lagrangian  the previous  iterations
can be  generalized in the following way
\footnote{
In eqs.~(\ref{algorithm}) we have implicitly assumed that all the fields
$\phi^\alpha$ are real. For charged fields, one can, obviously,
use the more conventional (and practical) complex notation. We did not do it
in order to avoid a tedious digression about the combinatorial factors.
}
\ba
{\cal L} & = &{1 \over 2} \phi^\alpha \phi^\beta
\tilde\Pi^{\alpha \beta}
+{1\over 6} \phi^\beta \phi^\gamma\phi^\delta
 {\cal O}^{\beta\gamma\delta}
\nonumber \\
b_{j,0}^\alpha&=&\chi_j^\alpha~~j=1,n \nonumber \\
 b_{j,0}^\alpha & = & 0~~j>n
  ~~~~~~~~n= number ~of~external~particle \nonumber \\
  ...& & ... \nonumber \\
b_{j,m}^\alpha&=&-{1\over 2}\Pi_{j,t}^{\alpha \beta}
 {\cal O}_{t,k,l}^{\beta\gamma\delta} \sum_{r+s=m-1}
b_{k,r}^\gamma b_{l,s}^\delta \label {itstep} \nonumber \\
{\cal A}_{p_1,...,p_n}&=&-{1\over 2}\sum_{s+r=n-2}b_{j,r}^\alpha
\tilde\Pi^{\alpha\beta}_{j,l}b_{l,s}^\beta
-{1\over 6}
\sum_{s+r+t=n-3}{\cal O}_{j,k,l}^{\beta\gamma\delta} b_{j,r}^\beta
b_{k,s}^\gamma b_{l,t}^\delta +
b_{j,n-2}^\alpha \chi_l^\beta \tilde\Pi^{\alpha\beta}_{j,l}
\nonumber\\
\label{algorithm}
\ea
where greek symbols denote lorentz or internal symmetry
indices, $\phi$ denotes a generic field of the theory (a sum
over distinct particles is understood),
$\chi_j^\alpha$ stand for a properly chosen source term (for instance ,
$\chi=1$ for a scalar, $\chi_\alpha=u_\alpha(p)$  for a Dirac
spinor with four components $\alpha$, etc...), $\Pi$ is the appropriate
propagator, $\tilde \Pi$ is
the inverse of $\Pi$ and ${\cal O}$ is the generalization
of the $D_{jkl}$ of the eq.~(\ref{eqsol}) which can depend on the spin,
the momenta and any other internal quantum  number.
In eq.~(\ref {algorithm}) we have included only trilinear interactions;
notice, however,
 that any operator with a higher dimension can always be
reduced to a set of trilinear and bilinear ones using a proper set
of auxiliary fields.

A final remark is in order here. In eqs.~(\ref{algorithm}) one has to take
properly into account particle statistic. Formally this can
be achieved introducing a set of Grassman variables $\epsilon_j$,
 $\epsilon_j\epsilon_k  + \epsilon_k \epsilon_j = 0$,
and setting $ u(p) \rightarrow \epsilon u(p)$ for the fermion sources
where now $u(p)$ is a vector of ordinary numbers. In practice this
means that
each term in the sums of eqs.~(\ref{algorithm}) enters with a relative sign
depending on the order of the  $b_j$.

 Finally one should keep in mind that a simplification closely parallel
to the one obtained in eq.~(\ref{simple}) do always occur.

The advantage of the iterative eqs.~(\ref{algorithm}) is that
they can now be easily
implemented  in a fortran code.

\subsection*{ALPHA, an ALgorithm to compute PHysical Amplitudes}
We have implemented a fortran code, ALPHA, based on the
solution of
eqs.~(\ref{algorithm}).
We have first studied a purely scalar theory which allows to check also
processes with an high number of particles.
We have then added fermions with a Yukawa type interaction and
verified
that the fermionic  statistic is properly accounted for. Finally
the complete electroweak Lagrangian, in the unitary gauge,
 has been added (at  present the  strong
interactions are absent).
  A very important test for the code  is provided by the gauge properties
of the theory.
The good behavior of the scattering
amplitudes, calculated by ALPHA, with the increasing center of mass
energies shows  that    delicate  gauge cancellations do occur.

 We have then compared
our  results  with known one and
 reproduced the cross sections for
 many processes with two particles in the final state and the rates for
the processes $e^+e^- \rightarrow W W Z$, $e^+e^- \rightarrow Z Z Z$
with a good accuracy.


We now want to add
 some technical details about  the code.
To stress again the simplicity of eqs.~(\ref{algorithm}) we remark
that the bulk of the code consists of less than 1000
 FORTRAN executable statements.
On top of this one has to add several {\it peripheral } subroutines,
for example subroutines to read the input, to return the Dirac $\gamma$
matrix, to perform four vector scalar products, etc.
All these {\it peripheral } subroutines have two common features:
they are completely trivial and they can be checked one by one in a completely
independent way and therefore they do not increase the  complexity of
debugging.

\begin{table}
\begin{center}
\begin{tabular}{|c|c|}
\hline
process & cpu time (seconds) \\ \hline
$e^+ e^- \rightarrow \mu^+\nu_\mu \tau^- \bar\nu_\tau$ & 128 \\ \hline
$e^+ e^- \rightarrow \mu^+\nu_\mu \tau^- \bar\nu_\tau
\gamma$ & 170\\ \hline
$e^+ e^- \rightarrow \mu^+\nu_\mu \tau^- \bar\nu_\tau\gamma\gamma$
& 395 \\ \hline
$e^+ e^- \rightarrow e^+\nu_e e^- \bar\nu_e$ & 180 \\ \hline
$e^+ e^- \rightarrow e^+\nu_e e^- \bar\nu_e\gamma$ & 323 \\ \hline
$e^+ e^- \rightarrow e^+\nu_e e^- \bar\nu_e\gamma\gamma$ & 877 \\ \hline
\end{tabular}
        \caption
{
Required CPU time for the computation (single precision)
 of 10000
matrix elements for  various processes.
All the Higgs couplings are set to zero.
The times are given in seconds
and the calculations have been performed with a DIGITAL machine
ALPHA 3000/600 with 64M of memory.
}
\label {cpu}
\end{center}
\end{table}

Another important feature, which is again linked to the simplicity
of eqs.~(\ref{algorithm}), is the efficiency achieved by the code.
 As a practical demonstration  of this fact we show
in table \ref{cpu}
 the CPU performance for
various processes. It is clearly seen that the increment in CPU time
is significantly smaller than the increment in the number of Feynman graphs
and in their complexity.
Furthermore ALPHA is only  a first prototype of the
algorithm and  we believe that a large room for optimization is left.

\subsection*{Emission of a hard photon in four fermion production at LEP
 I\negthinspace{I}}
In this section we will use our technique to compute the cross section
of four fermion production from $e^+~e^-$ collision at the LEP
I\negthinspace{I} center of mass energy and  when an additional hard photon
is found in the final state.

 One important  goal of the
LEP/SLC collider program has been achieved by  the precise measurements.
 After the first
phase which has revealed an  important and enlightening agreement for
most  \cite{disagreement,exp} of the
experimental data and the Standard Model predictions \cite{bardin},
 the measurement of  $W$-boson mass at LEP I\negthinspace{I} will
complete the pattern of precision tests.
 The importance of this goal  has been
pointed out by several authors \cite{veltman}.
The most promising  way to measure the charged boson mass  at LEP
I\negthinspace{I} is to make a precise  measurement of the cross
section near above the threshold of $W$-pair production.
 The two body phase space is
 proportional to the velocity $\beta$ of the two final $W$-boson;
thus the total cross section near the threshold   essentially  behaves
as $\beta$. The strong dependence of this parameter from the boson
mass means that the measurement of the total cross section can be seen
as a precise measurement of the $W$-mass.

Experimentally the $W$-bosons can be detected only through their decay
products. As a first  approximation the rate for each
process is given by the total $W$-pair cross section times the branching
ratios  of  W  into a pair of fermions, but the existence of a finite
width and of other non-resonant diagrams  introduce  sizable corrections.
Due to the experimental accuracy which is aimed
to at LEP I\negthinspace{I} the full  computation  of the process
\be
e^+~+e^- \longrightarrow~ 4~ fermions
\ee
is required\footnote {However, when proper kinematical cuts are applied,
there are semianalytical approximations which reproduce the correct result
very accurately} \cite{soffici}.
In the first column of table (\ref{tab1}) we show the results of the
program  ALPHA   for the process
\be
\label{processo}
e^+~+e^- \longrightarrow \mu^+~+~\nu_\mu~+~\tau^-~+~\bar\nu_\tau
\ee
near above the $W$-pairs production threshold, and in  the second column
 the set  of  the relevant input constants is shown.
We have applied a cut over the invariant mass of $\mu^+,$ $\nu_\mu$
and $\nu_\tau$ of $3$ GEV to eliminate the contribution from $\tau$
pair production with subsequent $\tau$ decay.

The value of the  $sin^2(\theta_W)$ and of the
$\alpha^{-1}_{QED}(2 M_W)$
given  in the table enter into the cross section
  through the tree  level formula of  the coupling describing the fermion
and gauge boson SM  interactions.
    \begin{table}
        \begin{center}
        \begin{tabular}{|c|c||c|c|} \hline
        $\sqrt{S}$ in GeV & cross sect. in pb & input parameter &
value
\\  \hline  \hline
	 150 & 0.00594 (3) & $ M_Z$ & 91.1888 $GeV$
\\
	 157 & 0.00937 (4)  &  $sin^2(\theta_W)$ & 0.23103
\\
	 160 & 0.04201 (14)  & $\alpha^{-1}_{QED}(2 M_W)$ & 128.07
\\
	 163&  0.09113 (26) & $M_W$ & 80.23 $GeV$
\\
	 170 & 0.17019 (52) & $\Gamma_W$ & 2.03367 $GeV$
\\
	 180 & 0.21166 (47) &$\Gamma_Z$ & 2.4974 $GeV$
\\
	 200 & 0.22891 (55) & $M_\tau$ & 1.7 $GeV$ \\
\hline
 \end{tabular}
        \end{center}
        \caption{ W-pair production rates (in pb) and the input
constants.
}
         \label{tab1}
\end{table}
Since the W-boson lifetime is finite, the cross section is only smoothly
suppressed below the $ S=4 M_W^2$ threshold; it increases up to the $ 200
  ~GeV$ region and after  decreases as $1/S$.
The results are in agreement within the montecarlo error, with
the existing   computations  for the process (\ref{processo}).

Beyond the {\it born
level}    contribution,
 the QED virtual and real corrections are the most significant.
Soft or collinear photons cannot be seen  by {\it real}
experimental detector, thus only a   cross section that includes the
emission of such photons  can be  compared with the experimental
results \cite{soffici}.

In the following we will not deal with the problem of
infrared and collinear QED correction to the process (\ref{processo})
and
we will present the results of the study of the exclusive process
\be
\label{procgam}
e^+~+e^- \longrightarrow \mu^+~+~\nu_\mu~+~\tau^-~+~\bar\nu_\tau ~
+ ~ \gamma.
\ee

The number of Feynman diagrams increases considerably
 and it is very difficult to handle the calculation
with the ordinary methods. The matrix element is a
function of eleven variables and also the numerical integration of such a big
number of variables is not easy.

To achieve the  numerical accuracy for each  total cross section shown
in table  (\ref{tab2}), a set of about  1.7 millions of events  has been
collected.
To perform the integral over the kinematical variables we have used
the package VEGA \cite{vega} to properly take into account
the peaking behavior of scattering matrix element.

 We  have introduced two cut-off to keep  the infrared
divergences under control and to provide  experimentally testable results:
\begin{itemize}
\item[-] $E_\gamma>1 Gev$ is the minimum photon energy required;
\item[-] $\theta>200~mrad$ ($cos(\theta)<0.98$)
 is the minimum angle allowed between the photon and each
  (both initial and final)   charged fermion in the process.
\end{itemize}

The table (\ref{tab2}) shows the total cross section at various
initial  energies. If an integrated luminosity of about $500~pb^{-1}$
is collected, few events  will pass the above cuts with a differential
distribution  as in figures (1) and (2).
It is clear from figure (1) that there is a  {\it
forbidden } region (i.e. too low rate) and a  region where some
events will  show up.
    \begin{table}
        \begin{center}
        \begin{tabular}{|c|c|c|} \hline
        $\sqrt{S}$ in GeV & cross sect. in fb & number of events (
$500 pb^{-1}$)
\\  \hline  \hline
	 150 &  0.464 (2) & 0.23
\\
	 157 &  1.300 (4) & 0.65
\\
	 160 & 2.874 (12) & 1.44
\\
	 163& 5.724 (17) &  2.35
\\
	 170 & 14.795 (56)  & 7.4
\\
	 180 & 20.638 (52) & 10.3
\\
	 200 & 24.070 (62) & 12.0
\\
\hline
 \end{tabular}
        \end{center}
        \caption{ $\tau^+~\nu_\tau~\mu^-~\bar\nu_\mu~\gamma$ production
rates: $E_\gamma>1~GeV$ and $\theta>200~mrad$ (see the
text). }
         \label{tab2}
\end{table}
\subsection*{Conclusion}
In this paper we have  suggested an iterative algorithm which
we  have found very simple and effective to compute scattering
amplitudes.

Differently from other approaches it never requires the explicit use of
Feynman graphs since  the truncated connected Green functions are
directly derived  by performing a numerical Legendre transform of the
effective Lagrangian. We also have shown that the algorithm
can be successfully  used to implement a fortran code,
that can  calculate very cumbersome  processes in a relatively short time.
To provide an example
of the performances
of our code  we
have computed  the cross sections  for some  relevant processes
of physical interest during the LEP I\negthinspace{I} phase,
namely the processes $e^+e^- \rightarrow \mu^+ \nu_\mu \tau^- \bar \nu_\tau$
and $e^+e^- \rightarrow \mu^+ \nu_\mu \tau^- \bar \nu_\tau\gamma$.

\subsection*{Acknowledgements}
We wish to thank G. Altarelli for   useful discussions.

\newpage
\subsection*{ {\bf Figure Captions}}
\noindent {\bf Figure 1:} differential cross section (fb/BIN) of
$e^+e^-\rightarrow \mu^+\nu_\mu\tau^-\bar\nu_\tau  \gamma$
 as function of $ 0.42<cos(\theta)<0.98 $ and $1~GeV<E_\gamma<19.5~Gev$
(for $E_{cm}=200~GeV$).\\
\noindent {\bf Figure 2:} differential cross section (fb/BIN) of
$e^+e^-\rightarrow \mu^+\nu_\mu\tau^-\bar\nu_\tau \gamma$
as function of $cos(\theta)$ (for $E_{cm}=200~GeV$).
$E_\gamma>1~GeV$ (see the text).


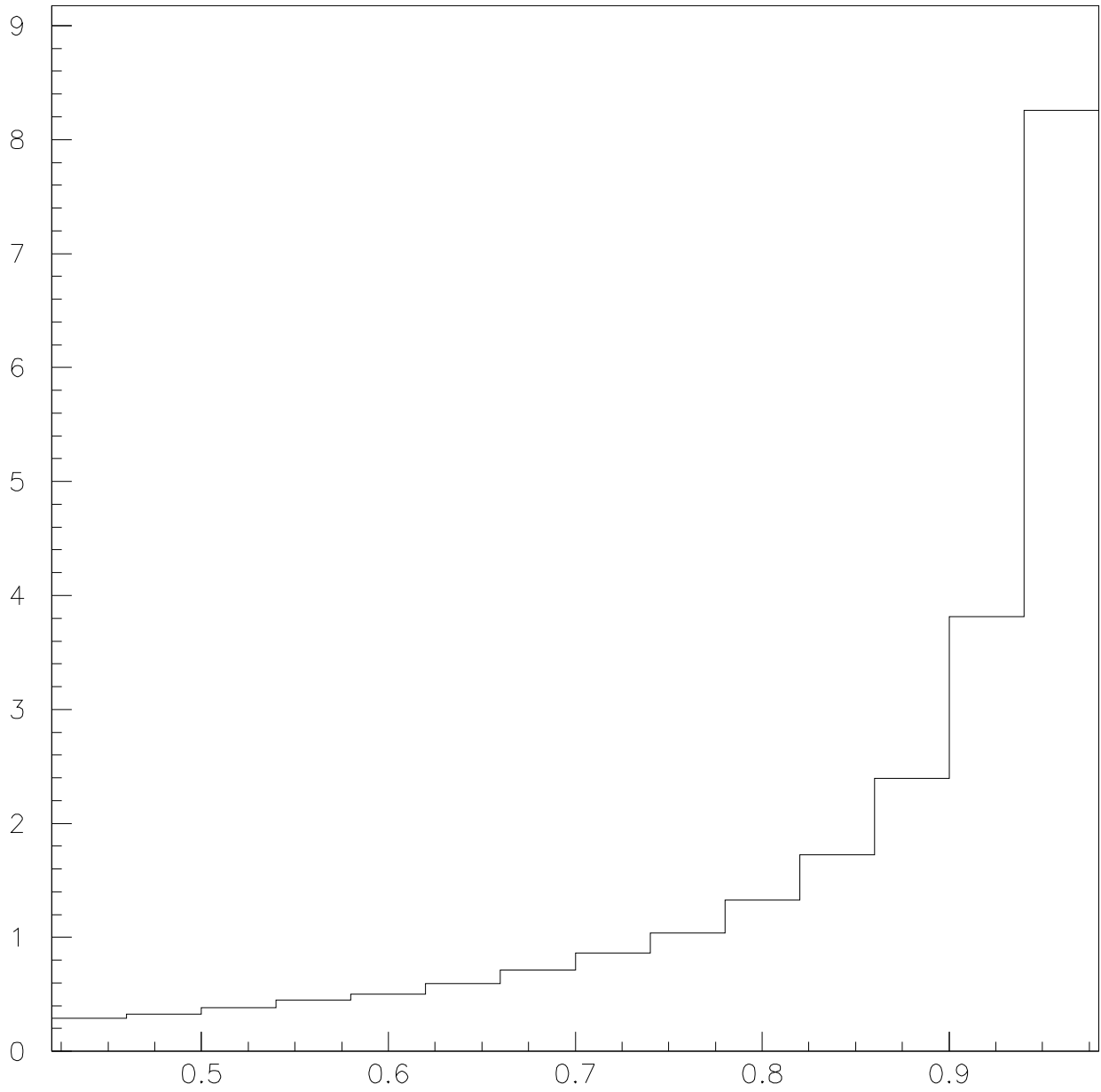
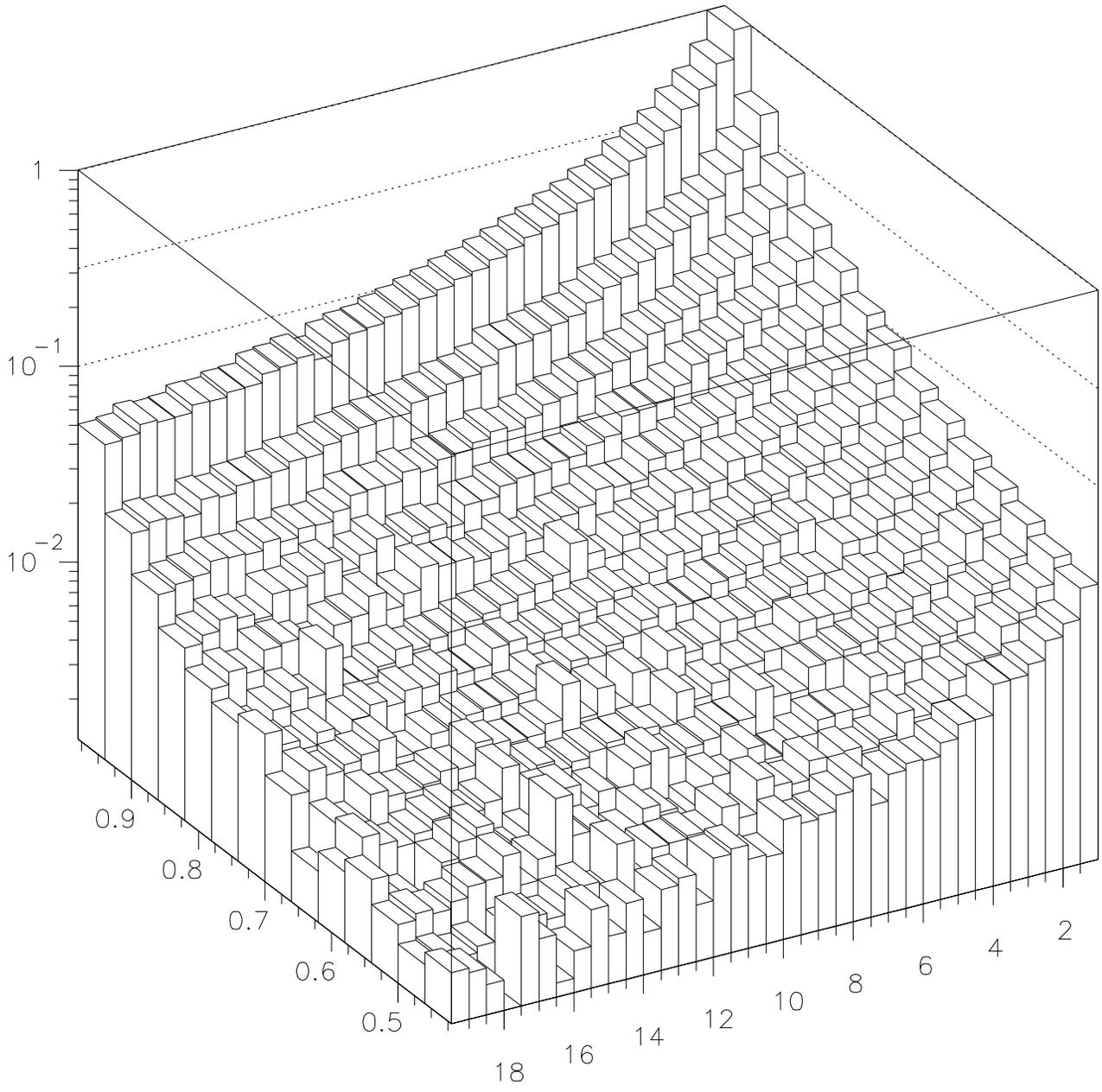
\end{document}